\documentclass[conference]{IEEEtran}
\IEEEoverridecommandlockouts
\usepackage{cite}
\usepackage{amsmath,amssymb,amsfonts}
\usepackage{algorithmic}
\usepackage{graphicx}
\usepackage{textcomp}
\usepackage{xcolor}
\usepackage{cite}
\usepackage{makecell}
\usepackage{multirow}
\usepackage{caption}
\usepackage{url}
\def\BibTeX{{\rm B\kern-.05em{\sc i\kern-.025em b}\kern-.08em
    T\kern-.1667em\lower.7ex\hbox{E}\kern-.125emX}}
\begin{document}

\title{Predicting batch queue job wait times for informed scheduling of urgent HPC workloads}

\author{\IEEEauthorblockN{Nick Brown}
\IEEEauthorblockA{\textit{EPCC} \\
\textit{University of Edinburgh}\\
47, Potterrow, Edinburgh \\
n.brown@epcc.ed.ac.uk}
\and
\IEEEauthorblockN{Gordon Gibb}
\IEEEauthorblockA{\textit{School of Physics and Astronomy} \\
\textit{University of Edinburgh}\\
Royal Observatory, Edinburgh}
\and
\IEEEauthorblockN{Evgenij Belikov }
\IEEEauthorblockA{\textit{EPCC} \\
\textit{University of Edinburgh}\\
47, Potterrow, Edinburgh}
\and
\IEEEauthorblockN{Rupert Nash}
\IEEEauthorblockA{\textit{EPCC} \\
\textit{University of Edinburgh}\\
47, Potterrow, Edinburgh}
}

\maketitle

\begin{abstract}
There is increasing interest in the use of HPC machines for urgent workloads to help tackle disasters as they unfold. Whilst batch queue systems are not ideal in supporting such workloads, many disadvantages can be worked around by accurately predicting when a waiting job will start to run. However there are numerous challenges in achieving such a prediction with high accuracy, not least because the queue's state can change rapidly and depend upon many factors. In this work we explore a novel machine learning approach for predicting queue wait times, hypothesising that such a model can capture the complex behaviour resulting from the queue policy and other interactions to generate accurate job start times. 

For ARCHER2 (HPE Cray EX), Cirrus (HPE 8600) and 4-cabinet (HPE Cray EX) we explore how different machine learning approaches and techniques improve the accuracy of our predictions, comparing against the estimation generated by Slurm. We demonstrate that our techniques deliver the most accurate predictions across our machines of interest, with the result of this work being the ability to predict job start times within one minute of the actual start time for around 65\% of jobs on ARCHER2 and 4-cabinet, and 76\% of jobs on Cirrus. When compared against what Slurm can deliver, this represents around 3.8 times better accuracy on ARCHER2 and 18 times better for Cirrus. Furthermore our approach can accurately predicting the start time for three quarters of all job within ten minutes of the actual start time on ARCHER2 and 4-cabinet, and for 90\% of jobs on Cirrus. Whilst the driver of this work has been to better facilitate placement of urgent workloads across HPC machines, the insights gained can be used to provide wider benefits to users and also enrich existing batch queue systems and inform policy too.
\end{abstract}


\section{Introduction}



The global pandemic has demonstrated the need to make urgent, accurate, decisions for complex problems. Each year there are many localised emergencies including wildfires, traffic accidents, disease, and extreme weather which, not only claim many lives and result in significant economic impact, but with the rise of global issue such as climate change, are becoming more prevalent. HPC has a long history of simulating disasters after the event, and recent technological advances have opened up the possibility of running simulations in real-time whilst disasters are unfolding. It is not just the increased computational power of such machines that unlocks such opportunities, but also the coupling with real-time data and improved technologies enabling interaction with simulations in real-time.

However, HPC machines are typically optimised for throughput and not latency of individual jobs. In short, the batch queue system means that there is an unbounded time in which simulation jobs will wait in the queue, and it is entirely useless for emergency responders to be waiting for insights from an HPC simulation job that is held in the queue whilst the forest is burning. Small scale urgent workloads previously addressed this issue by relying upon high priority queues or the ability to interrupt existing simulations \cite{spruce}, however this is not practicable for unpredictable and dynamic situations, where the amount of computing required can be large and vary significantly as time progresses \cite{technologies-urgent}. Consequently, leveraging several HPC machines and making informed choices about what workload to run where can provide more flexibility \cite{technologies-urgent}. 

A major part of this is understanding when a job will start to run on the supercomputer, which is a difficult undertaking as there are numerous factors which influence this. For example, job start time depends not just upon the configuration of the queue itself but also, at the time of submission, both the current queue state and jobs that will be submitted subsequently whilst our job of interest is still queued up. There are also aspects such as queuing policy, often determined by the HPC centre, which mean that the queues can be somewhat of a blackbox to users when they are trying to estimate how quickly their jobs will start to run. Consequently the use of machine learning based on historical batch queues data is an approach that has gained some traction, with the idea being that these models can capture the underlying patterns and use this to more accurately determine how long a job will wait in the queue. However existing machine learning models tend to be overly simplistic, limited by specific requirements or assumptions, or only target small-scale HPC or numbers of jobs. Whilst our driver is to obtain queue wait time predictions to optimise the placement of urgent real-time HPC workloads, better understanding the likely job queuing time has numerous additional benefits. These range from being able to better advise users what HPC machines to use and job parameters to select in their scripts, to better informing HPC centres so they can enhance queue policies and queue system settings to improve throughput.

In this paper we explore the use of machine learning to predict the queue wait time for jobs on HPC machines. Using data from real-world jobs submitted to production HPC machines, in Section \ref{sec:bg} we survey related work before describing the HPC machines used in this work and reporting performance of Slurm's built in estimator as a baseline. In Section \ref{sec:initial_ml} we describe our initial machine learning approach which, whilst simplistic, provides a limited degree of accuracy before building on this in Section \ref{sec:queue-model} to improve our prediction by incorporate the current state of the queue into our models. This is then further enhanced in Section \ref{sec:combination} by combining classification and regression to specialise the prediction of model regression, before briefly illustrating the use of our models to produce insights for users in Section \ref{sec:insights}. Lastly we draw conclusions in Section \ref{sec:conclusions} and discuss further work.

The contributions of this paper are:
\begin{itemize}
    \item Demonstration that the estimated time provided by Slurm's backfill plugin tool is inaccurate for real-world jobs and accuracy also varies significantly across machines.
    \item The ability to handle the highly unpredictable workload present on a supercomputer by developing a stochastic method which generates different random queue states that are still representative of machine usage patterns and used as inputs to the model.
    \item Illustration that whilst it is possible to generate moderately accurate predictions based on a simplistic model, by adopting a multi-model approach of classification and regression one is able to obtain more accurate results
    \item Presentation of job start time accuracy within a specific time frame of the actual start times, where the majority of related work instead presents accuracy using a metric such as mean standard error. Whilst such metrics are fine, our urgent use-case, and HPC users more generally, are likely more interested in how closely predictions will match to actual start times within a specific time frame.
\end{itemize}

\section{Background and related work}
\label{sec:bg}

The VESTEC marshalling and control system \cite{{technologies-urgent}} is a generic solution for running urgent, interactive workloads on HPC machines. Integrating use-cases ranging from wildfire fighting \cite{urgent1} to tracking mosquito-borne diseases \cite{mosquito}, these all represent highly dynamic workloads, often driven by the arrival of data from external sources and the requirement that such workloads must start to run as quickly as possible. To address these requirements the VESTEC marshalling and control system federates across multiple HPC machines, seamlessly distributing the execution of individual codes comprising the urgent workload across these machines and enabling users to interact with these in a location independent manner. Such federation is desirable because it provides a degree of resilience, if one specific machine fails then workload can be rescheduled, and also means that the urgent workload does not overwhelm one individual supercomputer. 
Disasters are thankfully relatively rare, so apart from the few highly specialised disaster tracking and relief organisations such DLR-GZS \cite{dlr-gzs}, it is not realistic to have dedicated resources set aside for these, but instead to be able to make use of existing very large scale supercomputers which normally run scientific or engineering simulations. The ultimate desire is to be able to leverage a large number of HPC machines, for instance all the tier-0 machines of Europe, when such a disaster unpredictably occurs. However this requires making informed decisions over job placement across these machines, and a major component of this is to be able to accurately predict how long jobs will wait in the queue on each individual machine before they start to run. 

\subsection{Queue wait time prediction}
Job scheduling algorithms typically rely on policies such as first come first served, shortest
job first, longest job first, or a job scoring methodology. However in order to ensure that the compute nodes of HPC machines remain filled and fairly allocated amongst users there are numerous additional complexities imposed at the system administration level which makes their operation far less transparent. Backfilling is one such example, where smaller jobs lower down the queue are prioritised to fill-up small numbers of available compute nodes which are not sufficient for the larger jobs higher up the queue. Not only does backfilling induce additional uncertainties when trying to understand how long a job will queue before starting to run, but furthermore there are numerous backfilling algorithms that can be selected making this even more opaque. Backfilling is just one example and, put simply, the queue systems of modern HPC machines are black-boxes, as such it is not a trivial task to predict how long jobs that are submitted to such queues will wait before they are allocated compute node(s) and start executing.

An early approach to job wait time prediction was first proposed in \cite{tsafrir}, which calculates the average time based on the previous jobs submitted by a user. Whilst this method is very simple, only needing to take the average of previous jobs, the accuracy was found to be lacking. By contrast \cite{Smith} followed a simulation approach where they simulated first come first served, shortest job first, and backfill scheduling algorithms. These were then used to predict the wait time for each application when that application is submitted to the scheduler. As part of this they also predicted the runtime of applications too, enabling a complete view of the state of the machine, and errors in job wait time predictions ranged between 5.01 and 996.67 minutes depending upon the workload being predicted and scheduling algorithm simulated.

Approaches exploiting historical queue data have become more common than queue wait time averaging or simulation. Supervised learning, where mathematical models are trained using the historical queue data is most popular and \cite{li} proposed an Instance Based Learning (IBL) approach to predict job start times based upon historical wait times. Whilst this work was early in the field of machine learning, with more limited techniques available to the authors and the number of jobs fairly small, their absolute average error ranging between 210.5 to 577.1 minutes which was promising for the time and acted as a foundation for further work. An example is \cite{queue-metascheduling}, which predicts the job queue waiting time by undertaking classification of similar jobs in the historical queue data. Their approach first predicts the wait time of a job using the K-Nearest Neighbour (KNN) technique and then undertakes classification using Support Vector Machine (SVM) among the classes of the jobs, with the probabilities given by the SVM used to provide a set of predicted wait times with probabilities. Whilst their experiments used the grid rather than an HPC machine, and they were aiming to predict within a window of 1 hour for the job start time, they were able to demonstrate correct categorisation of job start times between 77\% and 83\% of the time. By contrast in this work we are using real-world HPC machines, and after predictions that are as close as possible to the actual job start time rather than predicting within hour windows as that level of accuracy is not sufficient for the urgent use-case.

In \cite{park} the authors proposed a method of predicting queue wait times based on a hidden Markov model. They were interested in queue congestion, where the greater the congestion the longer the time before a job starts. In this work they represented queue congestion as an estimate of the state according to the degree of congestion for the queue waiting time expected at the time \emph{t}, with the objective being that they can then use their model to predict the queue waiting times at time \emph{t+1}. When comparing their prediction accuracy against those of the other methods, their results show that the proposed algorithm improves the prediction accuracy by up to 60\% although at only 10836 jobs their dataset is small.

By contrast to the supervised learning approaches detailed above, \cite{wang} studied the use of Reinforcement Learning (RL) to predict queue wait times, where a model is trained based upon rewarding desired behaviours and punishing undesired ones. In \cite{wang} the authors highlight the role of RL in handling the unknown amount of work in the queue, however their approach also requires prediction of the actual runtime of jobs before undertaking the job start time prediction. Undertaking runtime prediction is needed to accurately know the amount of work in the queue, but requires in-depth knowledge of each individual job and is not scalable to large systems with many different workloads. For instance in \cite{wang} the authors limited themselves to VASP only, and by contrast we aim for an approach which can be run on a snapshot of the machine executing a diverse workload without requiring such in-depth knowledge.

We summarise that previous work to predict queue wait times has demonstrated promise, but can be somewhat limited, based upon small scale datasets, or implies numerous assumptions that often do not generalise to real-world HPC machines \cite{adaptive-framework}. However all the papers surveyed in this section highlight the difficulty of predicting the job queue wait time and demonstrate this is complicated by a number of factors outside the control or knowledge of users. Often one does not know the full set of criteria that the scheduler uses to determine when jobs will run and there can be complicated relationships between these. Driven by our interest in urgent computing workloads \cite{urgent1}, we require the ability to quickly predict how long a job will queue for on a given HPC machine before running based because we require that urgent jobs start to run as soon as possible. This therefore means that we require the predicted start time to be reported in minutes and seconds, and it is important that the accuracy of such predictions are within a few minutes.

\subsection{Machines used in this work}
In the experiments detailed in this paper we work with historical data from the following three machines:
\begin{itemize}
    \item \textbf{ARCHER2}: An HPE Cray EX which is the UK national supercomputer and contains 5860 nodes, each with 128 CPU cores. At the time of writing ARCHER2 has been operational for four and a half months from December 2021 until mid April 2022, and the historical queue data we use in this paper comprises 314880 jobs in the standard queue and 73472 jobs in the short queue
    \item \textbf{Cirrus}: An HPE/SGI 8600 system with 280 nodes, each with 36 CPU cores. This is used by researchers across the UK but is more of a high-throughput system, targeting smaller jobs compared to ARCHER2. Our Cirrus historical data covers 15 months, from February 2021 until mid April 2022 and comprises 582200 (standard queue) jobs.
    \item \textbf{4-cabinet}: The preliminary 4-cabinet ARCHER2 system that was available before the full ARCHER2 system was commissioned. An HPE Cray EX, at 1000 nodes this was approximately a fifth the size of ARCHER2 and we use historical queue data over 9 months, from February 2021 to October 2021 which comprises 373560 (standard queue) jobs.
\end{itemize}

These three systems represent production HPC systems in use on a 24/7 basis, and provides a diverse set of system and job sizes to use for developing and evaluating our models. All systems run the Slurm \cite{slurm} queue system and we used the \emph{sacct} command to obtain the historical queue data.

It can seen that we report the number of jobs in the standard and short queue for ARCHER2, but just the standard queues for Cirrus and the 4-cabinet system. This is because, as we highlight in Section \ref{sec:initial_ml}, short queues are much easier to accurately predict job start times for compared with the standard queue. This is because jobs are always small, short running, and tend to start very quickly, in contrast to the standard queue which does not have these limitations and jobs can wait for considerable amounts of time \cite{archer2}. Therefore in the majority of this paper we focus on prediction for jobs in the standard queues across our machines as that is the major challenge for job start time prediction and also due to the limits of the short queue (e.g. jobs running for a maximum of 20 minutes), a workload of any complexity must use the standard queue.

\subsection{Slurm estimated time}
\label{sec:slurm_estimate}
The Slurm queue system \cite{slurm} provides its own job start prediction capabilities by providing an expected start time for jobs if Slurm is configured to use the backfill scheduling plugin. This prediction is only offered by Slurm once a job has been submitted, which is not quite suitable from our requirements in urgent computing, but nevertheless this estimated time acts as a baseline against which we can then compare the success of our machine learning approach in subsequent sections.

We tracked the lifetime of all jobs submitted on ARCHER2 and Cirrus over a two week period, using a script that continually polls the queue system for newly submitted jobs. Jobs are then stored and their details updated if required as time progresses, specifically amending the start time estimate if appropriate and once the job starts running our script compares Slurm's job start estimate(s) against the actual start time of that job.

Table \ref{tbl:slurm_prediction} reports the accuracy of the estimated start time provided by Slurm for the standard queues on both ARCHER2 and Cirrus. In this table we report the percentage of jobs whose start time estimates were accurate to within a specific time-frame of when the job actually started. Slurm updates the estimated start time for jobs if appropriate, and therefore there are two accuracy numbers reported for each machine in Table \ref{tbl:slurm_prediction}; the accuracy of the initial prediction made by Slurm when the job was submitted, and the best accuracy obtained across all the estimated start times for a job. For ARCHER2, on average, 83\% of jobs had multiple estimates provided by Slurm over their lifetime and 53\% of jobs had more than five updated estimates made. 

\begin{table}[h]
\begin{center}
    \begin{tabular}{|c|c|c|c|c|}
        \hline
        \multicolumn{1}{|c|}{\multirow{2}{*}{\shortstack{\textbf{Predictions accurate} \\ \textbf{within}}}} & \multicolumn{2}{c|}{\textbf{ARCHER2}} & \multicolumn{2}{c|}{\textbf{Cirrus}} \\ \cline{2-5}
        \multicolumn{1}{|c|}{} & \textbf{initial} & \textbf{best} & \textbf{initial} & \textbf{best} \\ \hline
        1 minute & 5.13\% & 16.55\% & 0.42\% & 4.12\%  \\
        5 minutes & 12.47\% & 23.51\% & 0.42\% & 4.26\%  \\
        10 minutes & 19.91\% & 30.99\% & 0.85\% & 4.40\% \\
        30 minutes & 41.31\% & 58.25\% & 2.69\% & 11.78\% \\
        1 hour & 58.40\% & 70.25\% & 4.40\% & 16.34\% \\
        2 hours & 69.91\% & 79.17\% & 8.38\% & 25.99\% \\
        6 hours & 81.47\% & 90.59\% & 30.11\% & 52.70\% \\
        12 hours & 90.45\% & 94.39\% & 50.85\% & 67.90\% \\
        24 hours & 95.39\% & 99.31\% & 77.98\% & 86.93\% \\
    \hline
    \end{tabular}
    \caption{Prediction accuracy of Slurm's backfill scheduling plugin for the standard queue on ARCHER2 and Cirrus. Both prediction accuracy for the initial estimate, when the job is first scheduled, and the best estimate over all updates is reported.}
\label{tbl:slurm_prediction}
\end{center}
\end{table}

From Table \ref{tbl:slurm_prediction} it can be seen that the estimates generated by Slurm are fairly inaccurate, especially for Cirrus. Irrespective of the machine in use the initial estimates are less accurate than the best estimate across all estimated start times an it can be seen that it is most common for predictions to be made that are accurate to an hour or more of the actual start time. The difference in Slurm's estimation accuracy is interesting between ARCHER2 and Cirrus, and this is because of the differences in usage model for the machines. Cirrus is a high-throughput system with many jobs requesting smaller numbers of nodes for a shorter amount of time. Consequently Slurm was overestimating the start time on Cirrus in 96\% of cases, whereas on ARCHER2, which follows a more traditional HPC system usage model, overestimation of the start time was in 60\% of cases. 

From Slurm's estimated start times we can conclude that the usage mode of the machine makes a significant impact to the accuracy of predictions generated. Slurm tends to overestimate, rather than underestimating, the start time and tends to be able to generate more accurate estimates for systems whose workload is more traditional HPC style. 

\section{Initial machine learning model}
\label{sec:initial_ml}
Previous work in \cite{quick-starters} and \cite{queue-metascheduling} demonstrated that K-Nearest Neighbour (KNN) \cite{knn} is a successful approach for generating queue wait time predictions, and a simple approach acting on the data was our starting point. KNN is a simple supervised machine learning algorithm for solving both classification and regression problems and works under the assumption that similar items exist in close proximity. Calculating the \emph{k} neighbours that are nearest to the feature of interest, these closest neighbouring values are then reduced to the overall prediction, often by taking the mean. An important configuration for KNN is what value of \emph{k} to use, i.e. the number of closest neighbours to each point that need to be considered. Based upon experimentation we found that \emph{k=10} was most appropriate, using the \emph{KNeighborsRegressor} from Sklearn \cite{sklearn} and the default \emph{minkowski} distance metric used to determine the nearest neighbours.

This simple KNN approach had two purposes, firstly to understand how even a very basic machine learning approach compares against Slurm's built in estimator, and secondly to act as a foundation for more complex machine learning approaches described in subsequent sections. Initially focusing on ARCHER2, based on the historical queue data we trained two regression KNN models. The first, \emph{basic}, only selects the number of nodes and wall time requested by the user as features for each job. The second model, \emph{temporal}, also includes the time and day of the week when the job was submitted as features and this enables us to understand the importance of when a job was submitted for accurately predicting job start time. Features for each job are normalised such that they all have approximately equal range, where we calculate the mean and standard deviation for each element for all jobs, and adjusting these as per Equation \ref{eq:normalisation} so that they are centred around zero, with a standard deviation of 1. 

\begin{equation}
f_i \to \frac{f_i - \left< f_i \right>}{\text{STD}(f_i)},
\label{eq:normalisation}
\end{equation} 

Throughout the experiments in this paper we select 80\% of jobs for training and 20\% for testing. However it was found that simply selecting every fifth job to be a test job resulted in a poor distribution of test data and artificially high predictions. This is because users can submit multiple similar jobs in one go, and therefore naively selecting one job in five for testing will mean that the models have likely seen very similar jobs previously. Instead to provide a fairer testing regime we select every fifth day of jobs to be a test job, for instance if all jobs on a Monday are selected as test jobs then the next test day will be a Saturday. This ensures that we have a wide range of test jobs across different hours and different days to test our predictions against sight unseen.

Table \ref{tbl:basic_knn} reports the results of prediction using our simple KNN models for ARCHER2. Accuracy is reported as the percentage of predictions that are made correctly within a specific time-frame of the actual job start time, with the smaller the difference the more accurate the model. We trained each model for both the ARCHER2 standard queue (314880 jobs) and short queue (73472 jobs) and it can be seen that the \emph{temporal} models, which are also provided with job submission date and time as features, result in increased accuracy of prediction compared against the \emph{basic} model. This demonstrates that there is a correlation between when a job is submitted to the HPC machine queue and how long it will wait for in the queue. Whilst such a statement will likely not surprise an experienced user of such machines from their own personal experience, it is still important to identify that this relationship exists in the data.

\begin{table}[h]
    \begin{tabular}{|c|c|c|c|c|}
        \hline
        \multicolumn{1}{|c|}{\multirow{2}{*}{\shortstack{\textbf{Predictions accurate} \\ \textbf{within}}}} & \multicolumn{2}{c|}{\textbf{Standard queue}} & \multicolumn{2}{c|}{\textbf{Short queue}} \\ \cline{2-5}
        \multicolumn{1}{|c|}{} & \textbf{basic} & \textbf{temporal} & \textbf{basic} & \textbf{temporal} \\ \hline
        1 minute & 9.27\% & 22.67\% & 33.53\% & 66.08\%  \\
        5 minutes & 28.79\% & 35.53\% & 46.46\% & 78.16\%  \\
        10 minutes & 39.53\% & 41.37\% & 50.47\% & 83.76\% \\
        30 minutes & 52.60\% & 53.32\% & 96.73\% & 92.21\% \\
        1 hour & 64.37\% & 61.85\% & 98.55\% & 96.18\% \\
        2 hours & 72.05\% & 68.68\% & 99.37\% & 97.50\% \\
        6 hours & 83.28\% & 80.70\% & 99.67\% & 99.23\% \\
        12 hours & 88.31\% & 87.40\% & 99.68\% & 99.67\% \\
        24 hours & 94.39\% & 92.68\% & 100.00\% & 100.00\% \\
    \hline
    \end{tabular}
    \caption{Prediction accuracy of simple KNN model on the queue data for ARCHER2 standard and short queues comparing basic model (the requested number of nodes and wall time only as features) and temporal model (also including job submission time and day).}
\label{tbl:basic_knn}
\end{table}

With the \emph{temporal} model for the standard queue in Table \ref{tbl:basic_knn}, when compared against the estimates generated by Slurm and reported in Table \ref{tbl:slurm_prediction}, it can be seen that this simple KNN approach improves the accuracy of predicted start times up to and including accuracies that fall within 10 minutes of the actual start time. Although this is at the cost of lesser accurate predictions, where for instance only 61\% of jobs in the ARCHER2 standard queue are correctly predicted to start within an hour of the actual start time, compared to 70\% for the estimation generated by Slurm. In contrast to the standard queue, predictions for the short queue are more accurate and there is less difference between the \emph{basic} and \emph{temporal} models. This is because there is much more uniformity to the short queue, small jobs with a short requested wall time being submitted to a set of reserved nodes, compared with the standard queue and as such this makes it much more predictable.

\begin{table}[h]
\begin{center}
    \begin{tabular}{|c|c|c|c|}
        \hline
        \makecell{\textbf{Predictions accurate} \\ \textbf{within}} & \textbf{ARCHER2} & \textbf{Cirrus} & \textbf{4-cabinet} \\
         \hline
        1 minute & 22.67\% & 51.48\% & 12.41\% \\
        5 minutes & 35.53\% & 61.20\% & 24.35\% \\
        10 minutes & 41.37\% & 65.28\% & 29.27\% \\
        30 minutes & 53.32\% & 75.38\% & 38.73\% \\
        1 hour & 61.85\% & 80.03\% & 44.87\% \\
        2 hours & 68.68\% & 88.47\% & 55.24\% \\
        6 hours & 80.70\% & 93.52\% & 71.18\% \\
        12 hours & 87.40\% & 96.43\% & 81.31\% \\
        24 hours & 92.68\% & 98.49\% & 89.34\% \\
    \hline
    \end{tabular}
    \caption{Prediction accuracy of simple KNN model on standard queue across our machines of interest}
\label{tbl:basic_knn_all}
\end{center}
\end{table}

We then trained our \emph{temporal} KNN regression model on the historical data from the standard queues of ARCHER2, Cirrus, and 4-cabinet. There is a separate model trained for each machine, and the results of using these trained models to test predicted job star times for 20\% of the data, sight unseen, are reported in Table \ref{tbl:basic_knn_all}. It is interesting to observe that predictions for job wait times on Cirrus using our simple model are considerably more accurate than those for ARCHER2 and 4-cabinet. This is in contrast with estimations made by Slurm and reported in Table \ref{tbl:slurm_prediction}, which were considerably less accurate for Cirrus than ARCHER2 and in all cases the simple KNN model predictions for Cirrus in Table \ref{tbl:slurm_prediction} beat what Slurm can provide. The high-throughput nature of Cirrus means that jobs on average tend to be smaller and faster running than ARCHER2 and 4-cabinet, thus starting more quickly. Consequently the model is biased towards predicting these pattern of jobs, and whilst not all jobs comply with this, enough do to make the predictions more accurate in this regime. Consequently on Cirrus pushing beyond this level of accuracy will be more challenging as to do so we must undertake accurate start time predictions for those jobs that do not conform with the most common machine usage pattern.

Even though the predictions in Table \ref{tbl:basic_knn_all} are far from perfect, given the simplicity of the model in use we were surprised at how well this performed. Compared against the accuracy of estimates provided by Slurm based on the \emph{best} accuracy reported in Table \ref{tbl:slurm_prediction}, it can seen for ARCHER2 that the KNN regression approach provides a greater number of predictions that fall very close (10 minutes or less) to the actual start time for ARCHER2 although there is a reduction for less accurate predictions. For Cirrus the KNN regression predictions are considerably more accurate than those provided by Slurm.

Whilst the prediction accuracy delivered by our simple model is encouraging, with our urgent workloads in mind this is not sufficient. Indeed more generally, users wanting to obtain a view of how long their application will queue on an HPC machine would expect to be able to obtain a reasonable accuracy within about 10 minutes or less.

\section{Queue snapshot machine learning approach}
\label{sec:queue-model}

\begin{table*}[!t]
\begin{center}
  \begin{tabular}{| l | l |}
  \hline
    Name & Description \\ \hline 
    nodes\_req & Number of nodes requested by the job \\
    req\_wtime & The requested wall time (hours) \\
    day & Day of the week (0-6) \\
    hour & Hour of the day (0-23) \\ \hline
    s\_q\_jobs & The number of queued jobs \\
    s\_q\_nodes & The total number of nodes requested by the queued jobs \\
    s\_q\_work$^*$ & The total work (nodes $\times$ wall time) of the queued jobs \\
    m\_q\_wait & The median time queued jobs have been waiting for to run (hours) \\
    d\_q\_nodes[0-7] & Histogram of the nodes requested by queued jobs  (8 values) \\
    d\_q\_work[0-7]$^*$ & Histogram of work requested by queued jobs (8 values) \\
    d\_q\_wait[0-7] & Histogram of wait times for queued jobs (8 values) \\ \hline
    s\_r\_jobs & The number of running jobs \\
    s\_r\_nodes & The total number of nodes requested by the running jobs \\
    s\_r\_work$^*$ & The total remaining work (nodes $\times$ remaining time) of the running jobs \\
    d\_r\_nodes[0-7] & Histogram of the nodes requested by running jobs  (8 values) \\
    d\_r\_work[0-7]$^*$ & Histogram of remaining work of running jobs (8 values) \\
    d\_r\_remain[0-7]$^*$ & Histogram of remaining times for running jobs (8 values) \\ \hline

  \end{tabular}
  \end{center}
\caption{The features used in our queue state aware machine learning model. Features with asterisks are calculated using actual wall times for historical data and for model testing. For prediction, randomly drawn wall times are used to calculate these. }
\label{table:feature_table}
\end{table*}

\begin{figure}[h]
  \centering
  \includegraphics[scale=0.57]{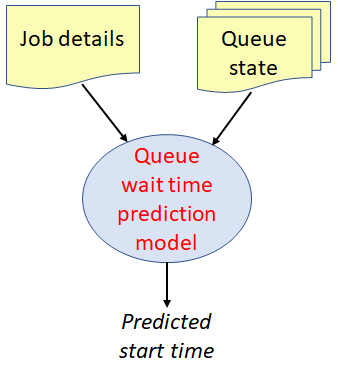}
  \caption{Illustration of model incorporating snapshot of queue for job start time training and prediction.}
  \label{fig:model_queue_simple}
\end{figure}

It was highlighted in Section \ref{sec:initial_ml} that for accuracy of prediction it is important to include as features when a job was submitted to the queue. This demonstrates there is a correlation between how long a job waits in the queue and when the user submitted it to the HPC machine. However, more generally, it is not the exact time and day when a job was submitted that is directly important, but instead the fact that this represents that the queue is in a specific state. For example at 10pm on a weekend the queue might be very quiet and submitted jobs will start to run quickly, whereas at 2pm on a Wednesday afternoon it is likely that many users are contending for the compute nodes and hence jobs will wait much longer in the queue. Whilst including the date and time of job submission as features improved the accuracy prediction of our simple model in Section \ref{sec:initial_ml}, it was our hypothesis that these are a rather crude way of representing the state of the queue and accuracy can be improved by providing the current state of the queue as features to our model when training and testing. This is illustrated in Figure \ref{fig:model_queue_simple} where queue state represents, at the time of job submission, all other jobs that are running or waiting to run on the HPC machine.

Consequently, for each job, we capture the state of the queued and running jobs at the time of submission and provide this to our model along with details of the job itself during training. Once trained, when making predictions for job start times using our test data, we take a snapshot of the running and queued jobs from the historically gathered data. When deploying our approach in the real-world this can be done in real-time via the appropriate Slurm commands and providing this as inputs to the model. 

Representing the state of queued and running jobs must be done in a manner that our models can easily consume. Specifically, providing values directly for every single queued or running job at the time of job submission would be cumbersome and liable to result in a very complex model. Instead, the current queue state is divided into queued and running jobs, and there are seven features representing the queued jobs and six representing the running jobs. These are summarised in Table \ref{table:feature_table}, and it can be seen that for both queued and running jobs there are three histograms. These histograms enable us to categorise the queue state into distinct bins that form a distribution, with these distributions representing the state of the queue.

Each histogram comprises eight bins and it is important to appropriately calculate the size and shape of each bins forming the histogram. To achieve this we aim for bins to roughly approximately to the same number of elements from a global perspective across all our jobs of interest in the training data. This is calculated by a script working through each submitted job in the training data and, based on the other jobs currently queued and running, will calculate the appropriate dimension of each histogram bin. It should be highlighted that whilst we aim for roughly each bin to hold the same number of elements from the global perspective, at the individual job level the size of each bin representing the current queue state often varies significantly and it is this which is providing the characterisation of the state.

Based on providing this enhanced queue state information we then retrained the KNN models and reran our experiments on the ARCHER2, Cirrus, and 4-cabinet systems for the standard queues. Using the same split and selection of training and test data as previously, the accuracy of our predictions when including the queue state are reported in Table \ref{tbl:queue_state_knn}. When comparing against prediction accuracy of our simple KNN model in Table \ref{tbl:basic_knn_all} it can be seen that providing the state of the queue as an input to the machine learning model tends to generally improve prediction accuracy, but this improvement is fairly limited. 

\begin{table}[htb]
\begin{center}
    \begin{tabular}{|c|c|c|c|}
        \hline
        \makecell{\textbf{Predictions accurate} \\ \textbf{within}} & \textbf{ARCHER2} & \textbf{Cirrus} & \textbf{4-cabinet} \\
         \hline
        1 minute & 28.52\% & 63.45\% & 18.43\% \\
        5 minutes & 37.09\% & 69.48\% & 24.54\% \\
        10 minutes & 38.81\% & 72.10\% & 32.88\% \\
        30 minutes & 59.34\% & 76.74\% & 39.28\% \\
        1 hour & 61.70\% & 79.89\% & 45.00\% \\
        2 hours & 70.20\% & 87.70\% & 62.28\% \\
        6 hours & 81.74\% & 91.60\% & 79.94\% \\
        12 hours & 88.18\% & 97.74\% & 91.44\% \\
        24 hours & 93.35\% & 98.73\% & 92.80\% \\
    \hline
    \end{tabular}
    \caption{Prediction accuracy when queue state is provided as an input to the KNN model}
\label{tbl:queue_state_knn}
\end{center}
\end{table}

We suspected that the limited improvement in predication accuracy reported in Table \ref{tbl:queue_state_knn} compared against the simple KNN model of Section \ref{sec:initial_ml} was because of uncertainties in the queue state. When a job is submitted we know exactly the number of other jobs already running and queued up waiting to run, and the number of compute nodes requested by the queued jobs and in-use by the running jobs. However we do not know exactly how long jobs will run for, and this was highlighted in \cite{wang}. Users provide a maximum wall time for their jobs, however when surveying the historical queue data we found that this maximum wall time tends to overestimate the actual job runtimes on average by around 8 times on ARCHER2 and 4-cabinet and 6 times on Cirrus. The most common source of these over estimations is where users select a default value, such as an hour or a day, as the maximum job wall time of their job.

Consequently when a job is submitted to the HPC machine there is uncertainty around how much work there is on the supercomputer in terms of how long running jobs will continue to run for and how long queued jobs will actually run for. The amount of work considerably impacts the start time of a job and approaches such as \cite{wang} look to address this by undertaking a prediction of runtime for currently running jobs. However predicting the runtime of running and queued jobs requires in-depth knowledge about those jobs, and instead it was our hypothesis that we could use this workload uncertainty as an advantage because it enables us to quantify the uncertainty on our predicted wait time. 

To address the uncertainty of work in the queue we adopt a stochastic approach where we train our KNN regression model on the actual wall times of jobs from the historical data, as these actual runtimes are known. When undertaking job start time predictions, at that point we only know the maximum specified wall time for jobs and therefore a large number of possible queue states are generated based on randomly chosen wall times for the queued and running jobs (as these are the unknowns). Each of these possible states are then run through our trained model as separate predictions and the resulting distribution of predicted job wait times is then used to determine the expected, mean, wait time. The error, standard deviation or similar, can also be generated to provide an estimate of accuracy. This confidence estimation is especially useful for the urgent computing case, as if the accuracy estimation is low then the VESTEC urgent computing system could submit the workload to multiple HPC machines and pick results from the job which ran through to completion first. 

\begin{figure}[htb]
  \centering
  \includegraphics[scale=0.31]{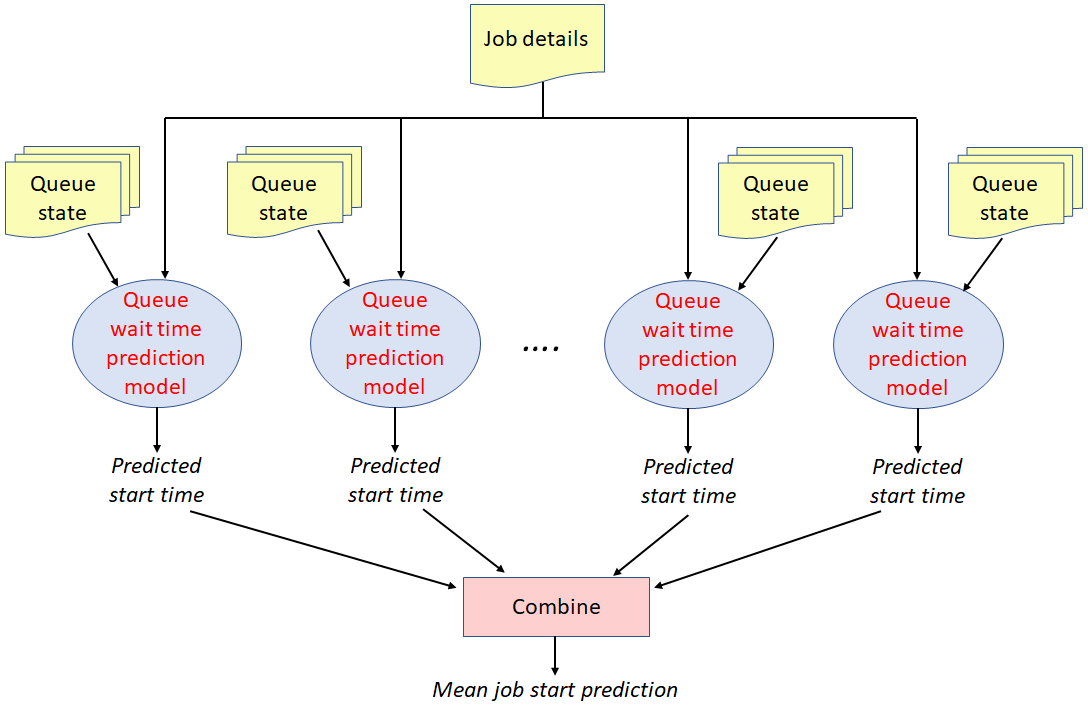}
  \caption{Illustration of stochastic approach operating over distinct randomly generated queue states}
  \label{fig:model_queue_refined}
\end{figure}

This stochastic approach for job start time prediction is illustrated in Figure \ref{fig:model_queue_refined}, where for clarity of presentation we only illustrate four distinct queue states being predicted by our trained model, although in reality we generate a hundred distinct queue states. It can be seen that the same job details are provided to each prediction, but each provided a separate queue state and generating its own distinct predicted start time. These predictions are then fed into the combination stage which generates the overall, mean, prediction and quantifies the uncertainty. 

\begin{figure}[h]
  \centering
  \includegraphics[scale=0.57]{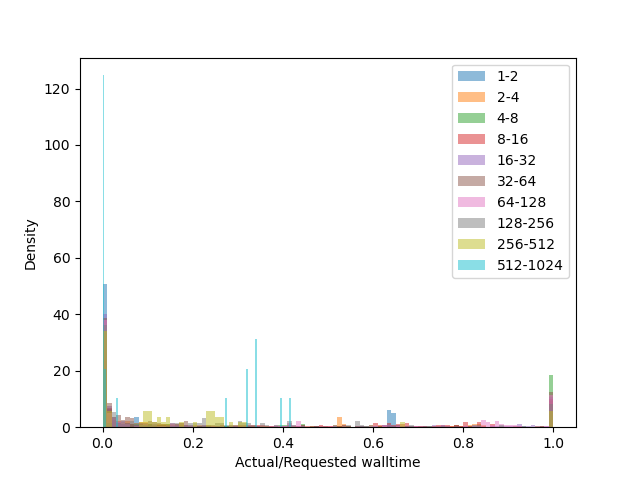}
  \caption{The distribution of actual to requested wall time for all jobs stratified by node count submitted to ARCHER2 in a single month}
  \label{fig:archer2_job_dist}
\end{figure}

For such a stochastic prediction method to work, we need to generate a set of random wall times for queued and running jobs as part of the hundred queue states. Ideally these will follow the same distribution of wall times as jobs previously submitted to the queue, and to achieve this, we consider the distribution of actual wall time to requested wall time for historical jobs. Figure \ref{fig:archer2_job_dist} illustrates our approach of distribution generation for ARCHER2 where we consider all jobs within certain node ranges and aim for such a random distribution of queue states for that specific machine to conform to this pattern. 

Based upon these distributions of the actual to requested wall times, we then generate a random number that obeys this distribution. This is achieved using the cumulative distribution function of the distribution chosen. Given the probability density function (PDF), in our case the histogram of Figure \ref{fig:archer2_job_dist}, we determine the cumulative distribution function (CDF) from Equation \ref{eq:cdf} where the CDF ranges from $[0,1]$.

\begin{equation}
\text{CDF}(x) = \int_{x_\text{min}}^{x} \text{PDF}(x')dx',
\label{eq:cdf}
\end{equation}

Consequently, based upon the CDF, if we pick a uniform random number, $x$, between $[0,1]$ we can then obtain a random number $y$ that follows the PDF by using Equation \ref{eq:pdf}. This approach enables us to generate random numbers that have the same distributions as the histogram shown in Figure \ref{fig:archer2_job_dist}, and using these random numbers we can determine random wall times by calculating \emph{y} times the requested wall time. Therefore whilst the runtime of each job of each queue state is random, these follow a realistic pattern given the jobs that are typically run on such a machine. 

\begin{equation}
 y = \text{CDF}^{-1}(x).
 \label{eq:pdf}
\end{equation}

From a code perspective once the raw job data has been cleaned and preprocessed into a usable state, we run a Python script which operates across the data and generates the \emph{CDF} wall time distributions. This data is then stored and used by the subsequent script which, for each job, constructs a list of the running and queued jobs in the queue at the job's submit time. For the 20\% of total jobs selected for testing, for each of these a hundred set of queue features are generated with the random wall times. Based on the predictions generated by our KNN model for each generated queue state we then take the mean of the $k$ nearest (based on a distance metric) vectors' actual wait times to be the predicted wait time. The prediction accuracy of this approach is reported in Table \ref{tbl:queue_state_refined_knn} for ARCHER2, Cirrus, and the ARCHER2 4-cabinet system. It can be seen that this stochastic queue generation approach improves prediction accuracy especially for ARCHER2 and the 4-cabinet system. Whilst even our simple KNN approach out-performed Slurm's estimations for the most accurate predictions, this is the first time where we outperform job start estimations generated by Slurm for all levels of accuracy across all machines.

\begin{table}[h]
\begin{center}
    \begin{tabular}{|c|c|c|c|}
        \hline
        \makecell{\textbf{Predictions accurate} \\ \textbf{within}} & \textbf{ARCHER2} & \textbf{Cirrus} & \textbf{4-cabinet} \\
         \hline
        1 minute & 41.70\% & 50.45\% & 20.41\% \\
        5 minutes & 54.16\% & 69.98\% & 27.16\% \\
        10 minutes & 61.17\% & 75.00\% & 45.00\% \\
        30 minutes & 69.09\% & 76.18\% & 51.48\% \\
        1 hour & 74.50\% & 79.65\% & 60.87\% \\
        2 hours & 76.12\% & 89.72\% & 78.57\% \\
        6 hours & 87.58\% & 92.36\% & 80.98\% \\
        12 hours & 91.73\% & 98.05\% & 93.96\% \\
        24 hours & 95.00\% & 99.67\% & 95.34\% \\
    \hline
    \end{tabular}
    \caption{Prediction accuracy of stochastic queue state approach, running our KNN regression model with one hundred randomly generated queue states for each job to determine the overall prediction.}
\label{tbl:queue_state_refined_knn}
\end{center}
\end{table}

\subsection{Boosted trees}
Until this point we have used a regression machine learning model based on K-Nearest Neighbours (KNN) as this was demonstrated to work well with queue predictions in \cite{queue-metascheduling} and \cite{quick-starters}. However KNN is a fairly simple approach and-so an important question was whether a more advanced technique would provide increased prediction accuracy. We explored the use of boosted trees \cite{boosted-trees} which model non-linear relationships in the data, and this is potentially advantageous as highlighted by \cite{wang} who themselves had some success with boosted trees in their work.

Otherwise known as gradient boosting, boosted trees rely on the concept of decision tree ensembles where a model consists of a set of classification or regression trees and features of the problem are split up amongst tree leaves. Each leaf holds a score associated with that feature and as one walks the tree, scores are combined which then form the basis of an overall prediction answer. A single tree is not sufficient for the level of accuracy required in practice, and so an ensemble of trees, where the model sums the prediction of multiple trees together, is used. As one trains a boosted trees model, the trees are built one at a time, with each new tree helping to correct the errors made by previously built trees. This is one of the factors that makes boosted trees so powerful and they have been used to solve many different machine learning challenges \cite{boosted1}\cite{boosted2}\cite{boosted3}. 

\begin{table}[h]
\begin{center}
    \begin{tabular}{|c|c|c|c|}
        \hline
        \makecell{\textbf{Predictions accurate} \\ \textbf{within}} & \textbf{ARCHER2} & \textbf{Cirrus} & \textbf{4-cabinet} \\
         \hline
        1 minute & 50.90\% & 65.86\% & 30.58\% \\
        5 minutes & 54.63\% & 74.44\% & 42.78\% \\
        10 minutes & 58.66\% & 77.55\% & 55.56\% \\
        30 minutes & 74.28\% & 82.76\% & 65.90\% \\
        1 hour & 79.74\% & 86.23\% & 73.70\% \\
        2 hours & 82.71\% & 88.33\% & 78.69\% \\
        6 hours & 93.71\% & 96.55\% & 85.78\% \\
        12 hours & 94.47\% & 98.52\% & 93.53\% \\
        24 hours & 97.50\% & 99.38\% & 97.23\% \\
    \hline
    \end{tabular}
    \caption{Prediction accuracy of stochastic predicted queue state boosted trees model across HPC machines of interest.}
\label{tbl:boosted_trees_first_attempt}
\end{center}
\end{table}

We used the XGBoost library \cite{xgboost} which is an open source software framework aiming to provide a scalable, portable and distributed gradient boosting library for Python and numerous other languages. Training our boosted trees model using the same approach for the stochastic queue state representation for the KNN model, results for this approach are reported in Table \ref{tbl:boosted_trees_first_attempt}. Whilst it can be seen that this generally provides more accuracy than the KNN approach in Table \ref{tbl:queue_state_refined_knn}, for instance it addresses the reduction in accuracy up to 1 minute for Cirrus seen in Table \ref{tbl:queue_state_refined_knn}, it is not a silver bullet. Still only around 55\% of jobs on ARCHER2 and 4-cabinet are correctly predicted to start within 10 minutes of the actual start time.

\section{Combining classification and regression}
\label{sec:combination}
In \cite{quick-starters} the authors improved the accuracy of their predictions by splitting their data on jobs that start within an hour, termed \emph{quick starters}, and longer waiting ones. They did this because the quick starters were commonplace and found to bias their models towards such predictions. In contrast, until this point we have been using regression models trained with 80\% of the historical data for each machine to generate a numeric queue wait time estimation. However, intuitively often users don't consider wait times to the exact second but instead within a specific bound, for instance whether the job will start within the next minute, 10 minutes, or hour. 

\begin{table*}[htp]
\begin{center}
    \begin{tabular}{|c|c|c|c|c|c|c|}
        \hline
        \multicolumn{1}{|c|}{\multirow{2}{*}{\textbf{Job start time category}}} & \multicolumn{2}{c|}{\textbf{ARCHER2}} & \multicolumn{2}{c|}{\textbf{Cirrus}}  & \multicolumn{2}{c|}{\textbf{4-cabinet}}\\ \cline{2-7}
        \multicolumn{1}{|c|}{} & \textbf{exact} & \textbf{relaxed} & \textbf{exact} & \textbf{relaxed} & \textbf{exact} & \textbf{relaxed} \\ \hline
        Immediately& 73.81\% & - & 89.69\% & - & 88.67\% & - \\
        Up to 1 minute & 73.60\% & 83.45\% & 88.38\% & 91.48\% & 68.45\% & 75.33\% \\
        Between 1 and 5 minutes & 63.98\% & 78.93\% & 75.48\% & 80.04\% & 43.26\% & 69.32\% \\
        Between 5 and 10 minutes & 61.51\% & 71.81\% & 62.79\% & 78.15\% & 61.95\% & 72.10\% \\
        Between 10 and 30 minutes & 75.90\% & 89.55\% & 65.54\% & 80.26\%& 71.93\% & 80.54\%  \\
        Between 30 minutes and 1 hour & 60.54\% & 82.20\% & 70.62\% & 84.16\%& 71.48\% & 74.52\%  \\
        Between 1 and 4 hours & 70.91\% & 74.77\% & 80.68\% & 84.55\% & 77.02\% & 82.26\% \\
        Over 4 hours & 80.26\% & 83.34\% & 87.03\% & 93.00\% & 77.55\% & 82.85\% \\
    \hline
    \end{tabular}
    \caption{Prediction accuracy of classification of jobs into start categories using boosted trees}
\label{tbl:queue_state_classification}
\end{center}
\end{table*}

Whilst the quick starters concept developed in \cite{quick-starters} was driven by grid computing rather than HPC, nevertheless when exploring the historical queue data on our HPC machines we found that a large proportion of jobs start within 10 seconds or less. Such quick job start times account for around 25\% of jobs on ARCHER2, 60\% of jobs on Cirrus, and 28\% of jobs on 4-cabinet. Consequently these frequent very short queue times bias our models during training for predicting shorter job queue times across the board. Therefore we modified our approach to the prediction of job start time by first defining categories of job start time and categorising jobs within these. This categorisation no longer involves generating an exact predicted time, as per regression, but instead for our model to determine which category of start time a job will reside in. We define the term \emph{immediate starters} which represent jobs that start within 10 seconds of being queued and first use a binary classification model to predict whether jobs are immediate starters or not. We focus our classification on the category with the most jobs in, for ARCHER2 and 4-cabinet the non-immediate starters category and for Cirrus immediate starters. Driving the grouping by this most numerous category because those jobs are more plentiful and hence easier to predict for, for instance with ARCHER2 categorisation is driven by those jobs predicted to be non-immediate starters, and every other job is assumed to be an immediate starter.

\begin{figure}[htb]
  \centering
  \includegraphics[scale=0.43]{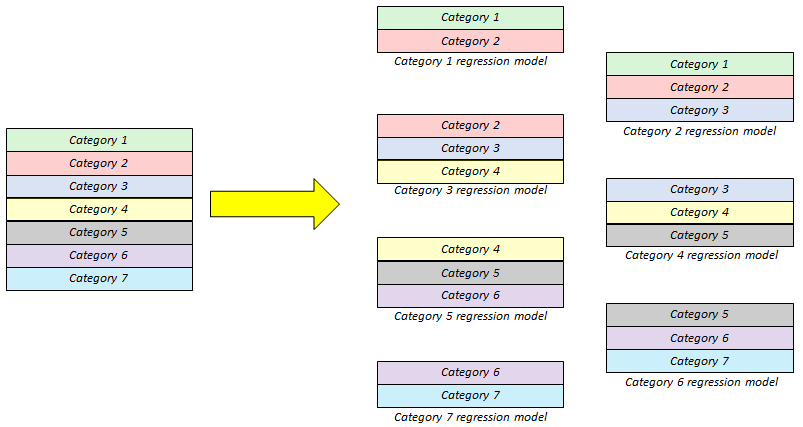}
  \caption{Illustration of categorising job start times and then for each category, red in this example, running the boosted trees model trained on that category and the one immediately preceding an following it.}
  \label{fig:combined_for_red}
\end{figure}

Those jobs not classified as immediate starters will then be fed to a subsequent model which categorises them into one of seven start categories, starting within a minute of being submitted to the queue, within 5 minutes, within 10 minutes, within 30 minutes, within 1 hour, within 4 hours, or stating over 4 hours after being submitted. For both classifications we found that using a boosted trees approach was most effective and Table \ref{tbl:queue_state_classification} reports accuracy of classification for our different job start categories. Whilst this classification follows the approach of \cite{quick-starters}, albeit we use boosted trees compared to \cite{quick-starters} who used SVM, in contrast it can be seen that we report both \emph{exact} and \emph{relaxed} accuracy. The \emph{exact} accuracy is the percentage of correct predictions made in that exact category, whereas the \emph{relaxed} accuracy is the percentage of predictions which are either correct or miss-predicted only in the category either side. The reason for this relaxed prediction was that we found it fairly common for some predictions to be close to the category boundary but the classifier is making a distinct choice, therefore whilst the prediction is not in the correct category it is close by. It can be seen from Table \ref{tbl:queue_state_classification} that the classification of jobs who are either immediate starters or starting within a minute of submission, is especially accurate. The accuracy is more variable for other categories although still tending to be fairly good for most.

\begin{figure*}[htb]
  \centering
  \includegraphics[scale=0.67]{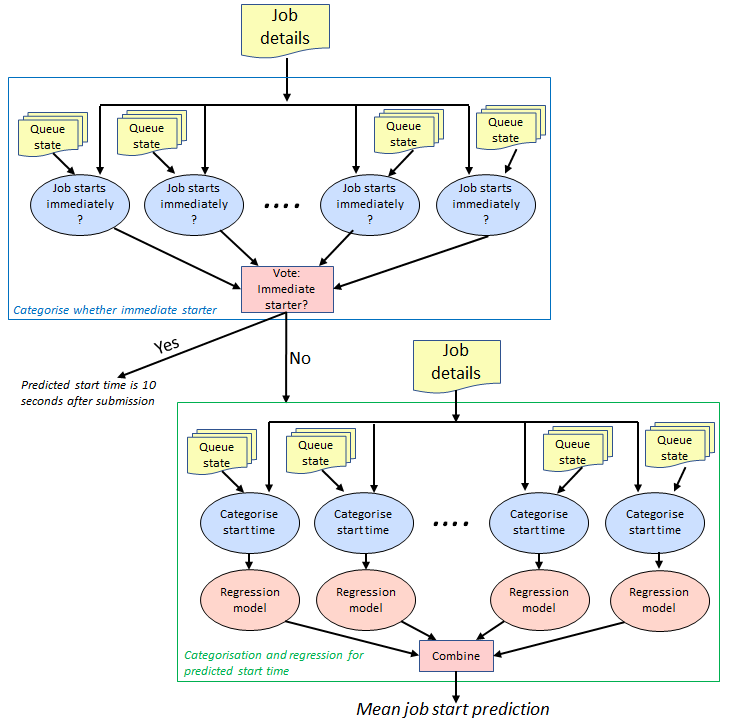}
  \caption{Illustration of overall flow for job start time prediction with combined classification and regression models}
  \label{fig:overall_flow}
\end{figure*}

In contrast to \cite{quick-starters}, we still want to obtain numerical start time predictions and Figure \ref{fig:overall_flow} illustrates the overall flow of our modified approach when predicting the start time of a job. The binary classification of whether the job will start immediately or not is first undertaken for all hundred stochastic queue states and the dominant decision, whether it is immediate or not, is selected. If it is selected as an immediate starter then the predicted job start time is set to be 10 seconds after the queue time. Otherwise the job is fed into instances of the classification and regression models, each running with the one hundred distinct queue states. For each of these queue states, once the job start category has been determined then the appropriate regression model for that category is selected and used generate the predicted start time. All hundred predicted start times are then combined, with the mean prediction taken as the overall start time prediction as described in Section \ref{sec:queue-model}.

The regression model for each category has been trained on data from that category and the categories either side of it. The idea is that, as per Table \ref{tbl:queue_state_classification}, if a job is going to be miss-categorised then it is most likely to have been so into one of the categories either side of the correct one. Consequently by training a regression model with three categories of data it provides the opportunity for the these miss-predictions to \emph{snap back} to within the correct category. 

This is illustrated in Figure \ref{fig:combined_for_red}, where those jobs not categorised as immediate starters are categorised as starting within one of seven timing categories on the left of the figure. Boosted trees regression models are trained for each category using data from the category itself along with the categories either side. The appropriate pre-trained model is then selected for undertaking predictions as illustrated in the overall flow depicted in Figure \ref{fig:overall_flow}. Table \ref{tbl:final_accuracy} reports overall accuracy for this approach, where it can be seen that this results in considerably improved prediction accuracy compared to the models in Section \ref{sec:queue-model}. Our combined classification and regression approach correctly predicts jobs starting within 1 minute for 63\% of jobs on ARCHER2, 76\% on Cirrus, and 66\% on 4-cabinet compared with Slurm's estimator reported in Table \ref{tbl:slurm_prediction} that only accurately predicts jobs starting within a minute at-best 16\% of the time on ARCHER2 and 4\% of the time on Cirrus. From Table \ref{tbl:final_accuracy} it can also be seen that we are reporting three quarters of all job start times accurate within 10 minutes on ARCHER2 and 4-cabinet, and 90\% on Cirrus, compared with Slurm's estimator that predicts accurately within 10 minutes only a 30\% of the time for ARCHER2 and 4\% of the time for Cirrus.

\begin{table}[h]
\begin{center}
    \begin{tabular}{|c|c|c|c|}
        \hline
        \makecell{\textbf{Predictions accurate} \\ \textbf{within}} & \textbf{ARCHER2} & \textbf{Cirrus} & \textbf{4-cabinet} \\
         \hline
        1 minute & 63.49\% & 76.87\% & 66.25\% \\
        5 minutes & 71.46\% & 88.31\% & 71.89\% \\
        10 minutes & 75.67\% & 89.69\% & 75.29\% \\
        30 minutes & 81.93\% & 92.11\% & 80.37\% \\
        1 hour & 85.17\% & 93.55\% & 83.19\% \\
        2 hours & 89.11\% & 95.39\% & 86.63\% \\
        6 hours & 95.87\% & 98.65\% & 94.74\% \\
        12 hours & 98.99\% & 99.68\% & 99.79\% \\
        24 hours & 99.93\% & 100.00\% & 100.00\% \\
    \hline
    \end{tabular}
    \caption{Prediction accuracy of combination of classification and regression boosted trees models}
\label{tbl:final_accuracy}
\end{center}
\end{table}

\subsection{Model runtime}
\begin{table*}[htb]
\begin{center}
    \begin{tabular}{|c|c|c|c|c|}
        \hline
        \textbf{Activity} & \textbf{Type} & \makecell{\textbf{ARCHER2} \\ \textbf{(seconds)}} & \makecell{\textbf{Cirrus} \\ \textbf{(seconds)}} & \makecell{\textbf{4-cabinet} \\ \textbf{(seconds)}} \\
         \hline
        Cumulative Distribution Function (CDF) & Training & 264 & 420 & 310\\ 
        Histogram bin identification & Training & 6981 & 11298 & 7256\\
        Split training and test data including binning & Training & 4322 & 6223 & 5098\\ 
        Generating random queue test states & Training & 7650 & 14890 & 8442\\ 
        Train Section \ref{sec:queue-model} stochastic queue KNN model & Training & 20884 & 70667 & 33410 \\ 
        Train Section \ref{sec:combination} classification and regression boosted trees models & Training & 13574 & 34589 & 15911 \\ \hline
        Single job prediction for Section \ref{sec:queue-model} stochastic queue KNN model & Prediction & 0.88 & 0.89 & 0.89 \\
        Single job prediction for Section \ref{sec:combination} classification and regression boosted trees models & Prediction & 0.11 & 0.18 & 0.10\\
    \hline
    \end{tabular}
    \caption{Runtime of model training and prediction activities on a 26-core Intel Xeon Platinum (Skylake) 8170 CPU}
\label{tbl:model_time}
\end{center}
\end{table*}
In this paper we have mainly focused on the prediction accuracy of our machine learning models. However such models must be realistic to run for jobs, especially with our focus of undertaking predictions of urgent workloads as part of the VESTEC system which needs to rapidly make decisions around job placement across numerous supercomputers. Consequently the machine runtime of models is also important, especially when undertaking job start predictions. We ran all our machine learning scripts on a 26-core Intel Xeon Platinum (Skylake) 8170 CPU with the runtime in seconds for different aspects reported in Table \ref{tbl:model_time} for each machine. With the exception of CDF generation and histogram bin identification, all codes were threaded across all 26 cores and it can be seen that model training is by far the most time consuming activity, although it only needs to be performed once per machine, whereas queue wait time prediction for a single job is less than a second. The boosted trees classification and regression models described in Section \ref{sec:combination} ran faster than the KNN model described in Section \ref{sec:queue-model}, training in considerably less time and also generating predictions in less time too. This was unexpected given the more advanced nature of boosted trees compared to KNN and fact that the boosted trees models undertakes classification and regression. Irrespective, for job wait time prediction the runtimes are small and we highlight that our approach is realistic to be used as in a semi real-time fashion, returning results in approximately a tenth of second for each machine when using our most accurate prediction approach.

\section{User insights gained from models}
\label{sec:insights}
\begin{figure}[h]
  \centering
  \includegraphics[scale=0.32]{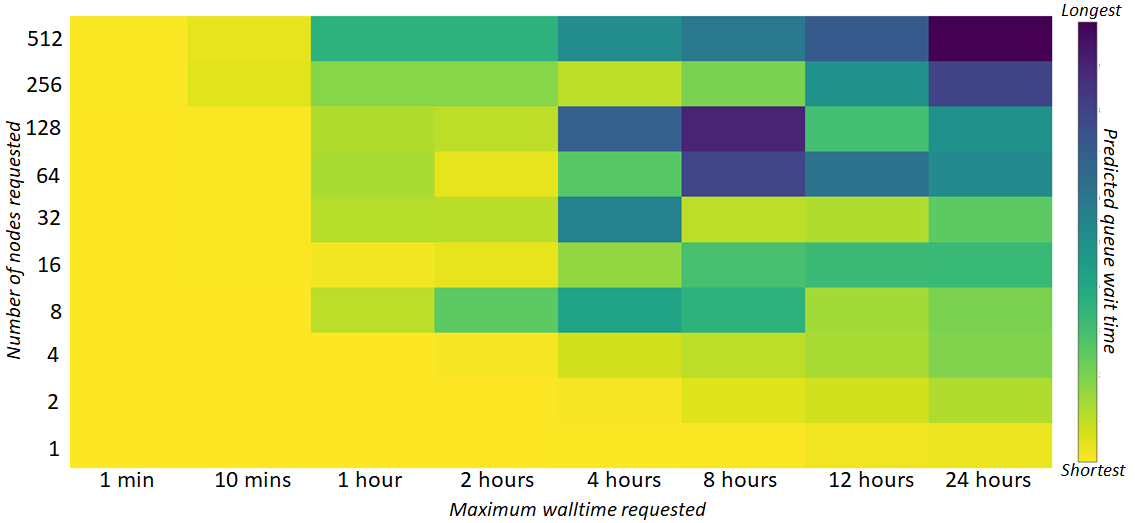}
  \caption{Predicting job queue wait time on ARCHER2 for different numbers of nodes and maximum wall times requested the the queue representing a standard Tuesday morning at 11am}
  \label{fig:insights}
\end{figure}

Our main objective in this paper has been to develop a model that can accurately predict job wait times for urgent workloads, enabling our VESTEC system to make informed choices around workload placement. However these models can be used more widely by users to help understand optimal job configurations when submitting to the queue. For instance answering questions such as whether changing the number of nodes will impact the overall queue wait time or the maximum wall time. Based on the models developed in Sections \ref{sec:queue-model} and \ref{sec:combination}, we undertook a number of predictions for the queue state of ARCHER2 on a standard Tuesday morning at 11am, for different number of nodes and maximum requested wall times. These predictions are illustrated in the heatmap of Figure \ref{fig:insights}, where users can obtain specific insights. For example it can be seen that if the user was after running over 16 nodes then if possible they should set a maximum requested wall time of 2 hours or lower, because from 4 hours requested wall time onwards the queue wait time increases sharply. If the user was requesting 32 nodes, then it can be seen that they should avoid requesting 4 hours maximum wall time as this is predicted to result in a much longer queue wait time than, for instance, requesting 8 or 12 hours maximum wall time with 32 nodes. Whilst these are simple examples, they illustrates how our prediction models can be used to provide insights to users around how changing the number of requested resources will impact their queue wait time and hence make more informed choices around job configuration.

\section{Conclusions}
\label{sec:conclusions}

In this paper we have explored the use of machine learning to predict job start times on three production HPC machines that represent a diverse size of machine and usage model types. Beginning with job start time estimates provided by Slurm as a baseline, we then explored the accuracy of a simple KNN model. Building upon this simple KNN model we explored how to provide the queue state as an input to our models, however this was found to be further complicated by uncertainties in the overall amount of work in the queue at job submission time as only maximum wall times are provided which can vary significantly from the actual runtime. Consequently we devised a stochastic approach which generates one hundred different queue workload states for each job and whilst these are random, being based on the distribution of wall times from jobs previously submitted they are still a realistic representation.

After exploring the improved prediction that our stochastic approach provides, both with KNN and boosted trees techniques, we then further developed this into a multi-stage approach which combines classification and regression. By adopting this approach we demonstrated significantly improved prediction accuracy for job start time, predicting within 1 minute for around 65\% of jobs on ARCHER2 and 4-cabinet, and 76\% of jobs on Cirrus, as well as accurately predicting three quarters of all job start times within 10 minutes on ARCHER2 and 4-cabinet, and 90\% of jobs on Cirrus. This represents a 3.8 times more accurate prediction for ARCHER2 and 18 times more accurate for Cirrus when compared to Slurm's estimations within a 1 minute accuracy window. When considering a 10 minute window our approach is 2.2 times more accurate for ARCHER2 and 20 times more accurate for Cirrus than Slurm's estimations.

The models we have developed can also be used to provide enhanced insights to users around when they can expect their jobs to run. In Section \ref{sec:insights} we provided an example of this around changing the number of nodes and requested wall time, with the job submission time fixed to be an average Tuesday at 11am, but it is also possible to vary the queue submission time and explore how submitting jobs at different times might reduce the overall wait times and this would be interesting to explore. Our approach could be incorporated into a tool that users can use to dynamically explore the most appropriate parameters for their jobs to optimise configurations, as well as potentially enhancing the Slurm queue system to provide more accurate job start predictions. 

To better improve the accuracy of our models, when binning the queue state we could consider the distribution of actual to requested wall times within certain wall time ranges, for instance for jobs with requested wall times less than one hour. This would provide a more accurate possible distribution for each given job, although for some edge cases there may be few jobs to draw a distribution from which would impact accuracy. Furthermore whilst our start time predictions take into account the existing jobs currently running and queued at job submission, we do not consider additional jobs that might be submitted after the job of interest has been queued but is still waiting to run. These subsequent jobs could impact the job start time and it would be possible to generate some stochastic representation of the likelihood and dimensions of such jobs, providing these as part of our queue state model inputs which could further improve prediction accuracy.

We conclude that our approach significantly improves upon the prediction accuracy of Slurm's estimator, and in contrast to existing machine learning techniques for predicting job wait time on HPC machines we are able to generate numerical job start times that tend to fall within one and ten minutes of actual job start times across our machines of interest. Our approach has been incorporated into the VESTEC urgent computing system to accurately predict job queue wait times across many different machines to ensure suitable job placement, and is also of benefit more widely to users and system administrators of HPC machines.

\section*{Acknowledgment}

The research leading to these results has received funding from the Horizon 2020 Programme under grant agreement No.\ 800904. This work used the Cirrus UK National Tier-2 HPC Service at EPCC (http://www.cirrus.ac.uk) funded by the University of Edinburgh and EPSRC (EP/P020267/1). This work used the ARCHER2 UK National Supercomputing Service (https://www.archer2.ac.uk).

\end{document}